\documentclass[12pt, preprint]{aastex}
  \begin{document}
\title{Robust Models for Phase Shifts in Accreting Binary Stars}
\author{Andrew G. Cantrell and Charles D. Bailyn}
\affil{Department of Astronomy, Yale University. PO box 208101, New Haven, CT, 06520}
\email{cantrell@astro.yale.edu, bailyn@astro.yale.edu}

\begin{abstract}

Radial velocity studies of accreting binary stars commonly use accretion disk emission lines to determine the radial velocity of the primary star and therefore the mass ratio.  These emission line radial velocity curves are often shifted in phase from the expected motion of the primary.  These phase shifts cast doubt on the use of disk emission lines in the determination of mass ratios.  We present a systematic study of phase shifts, using data from the literature to distinguish between possible explanations of the phase shift.  We find that one widely adopted class of models is contradicted by observations (section 2).  We present a generalized form of another class of models, which we call ``measurement offset models."  We show that these models are quantitatively consistent with existing data (figures 2 and 3, and the discussion in section 4.4).  We consider the implications of adopting measurement offset models, for both disk structure and determination of binary parameters.  Specifically, we describe in section 6 how measurement offset models may be used improve determinations of the mass ratio based on disk emission lines.  This could be a valuable new tool in determining masses of important astrophysical objects such as accreting neutron stars and black holes.

\end{abstract}


\maketitle
\section{Introduction}

Dynamical indicators have long been of central importance to the study of binary star systems, as measures of binary system parameters and therefore of the masses of the objects.  The masses are of general interest for constraining stellar models, and are of particular interest in compact accreting binaries.  These systems generally consist of an essentially normal star which is losing mass to a compact object.  The compact object is a  white dwarf in the cataclysmic variables (CVs) and a neutron star or black hole in low-mass X-ray binaries (LMXBs).  The nature of the primary, constrained by its mass, is of fundamental importance to a system's observed properties, allowing direct comparison of different types of compact objects.  In this paper, we consider the use of disk emission lines as dynamical indicators in accreting binaries, and describe a new method by which more trustworthy mass determinations may be made from these lines.

The robust determination of masses is of particular interest in objects containing black holes and neutron stars, since mass is the most reliable way of distinguishing between the two.  Suggestive evidence for the existence of event horizons in dynamically confirmed black hole systems was given by the observation (Garcia et al. 2001) that these systems have quiescent X-ray luminosities two orders of magnitudes smaller than their neutron star counterparts.  The black hole and neutron star mass distributions are also of interest as a constraint on the endpoint of stellar evolution (Bailyn et al. 1998, Brown et al. 2000 and references therein).  

Most compact accreting binary systems show strong disk emission lines, which potentially contain a wealth of dynamical information.  The lines are often double-peaked, providing compelling evidence for the existence of disks in these systems; the profiles reflect the dynamical structure of the disk.    The accretion disk should track the motion of the primary, and its emission lines are generally bright, making radial velocity (RV) measurements of the primary easy in principle.   However, the use of emission lines to measure the motion of the primary is hampered by the widespread observation of ``phase shifts," in which the  RV curve of the disk emission lines is not in antiphase with the RV curve of the secondary, or does not have the correct phasing relative to eclipses.  These phase shifts are commonly $20^\circ$ or more, and in the most extreme cases are near $90^\circ$.  Very large phase shifts are a defining characteristic of SW Sex-type objects (e.g. Dhillon, Marsh, \& Jones 1997), and substantial phase shifts have been observed in many others, including LMXBs, and CVs of many types; specific systems with references are given in table 1, which shows that phase shifts have been observed in a large majority of accreting binaries in which the necessary data exists. 

The existence of phase shifts shows definitively that RVs obtained from disk emission lines do not reflect the actual motion of the primary, casting doubt on their use in the determination of binary orbital parameters.   Since disk emission line RV curves apparently measure something other than the true radial velocity of the primary, we cannot assume that the semiamplitude of the disk  emission line RV curve is the same as the semiamplitude of the primary's true RV curve.  Though the latter is needed for a determination of the mass ratio from RVs, the former is all that can be determined directly from disk emission line RVs; it is therefore doubtful that a direct use of these emission lines gives a robust measurement of the mass ratio. 

In order to make robust measurements of the mass ratio using disk emission lines, we must understand the relationship between the semiamplitude of the disk emission line RV curve and the semiamplitude of the primary's RV curve.  Specifically, we would like to know when, if ever, the two are the same, and whether the disk emission lines are a more reliable measure of the primary's RV semiamplitude when small phase shifts are observed.  Ultimately, we would like to have a detailed understanding of the phenomenology of the phase shift, to allow a robust determination of the true RV semiamplitude of the primary, given RV data from disk emission lines.  This would allow disk emission lines to be used for the robust determination of mass ratios.

The relationship between RV curves obtained from disk emission lines and the true RV of the primary is model-dependent, and therefore any analysis of disk emission lines requires understanding the phase shift's origin.  In this paper, we describe two scenarios which have appeared in the literature as explanations of the phase shift, and which make different predictions about the relationship between the radial velocity semiamplitude of disk emission lines and the true RV semiamplitude of the primary.  We test predictions of each scenario against existing data, and show that one of the two is contradicted by a range of observations, while the other is well-supported by existing data and naturally explains some previously puzzling observations.  We discuss the implications of our adopted scenario in terms of the relationship between the RV semiamplitude given by disk emission lines and the true RV semiamplitude of the primary; we also discuss the implications of our work for future studies of disk structure.  

In Section 2, we summarize the two common explanations for the phase shift, and argue that at least one of these two must be invoked by \it any \rm possible model of the phase shift.  We present evidence contradicting the more widely adopted of the two models, and discuss basic principles of the alternative, which we call ``measurement offset models".  In Section 3, we discuss the basic characteristics of measurement offset models, and the conclusions that can be drawn from them; in section 4, we present a compilation of data on phase shifts and use it, together with the work of section 3, to test measurement offset models.  In section 5, we give special consideration to V2107 Oph and V838 Her, which are unique in showing negative phase shifts; we show that their unusual characteristics are well-explained by measurement offset models.  Section 6 describes possible future work directed both at better constraining models of disk structure and developing a technique for the robust measurement of mass ratios using disk emission line radial velocities.  Our conclusions are summarized in section 7.

\section{Two Explanations for the Phase Shift}

The presence of phase shifts in disk emission line radial velocity data shows that the radial velocity measured using these lines is not equal to the true velocity of the primary.  There are two possible explanations for this: either the line-emitting region does not share the motion of the primary, or the velocity of the line-emitting region is not accurately reflected by RV measurements.  At least one of these two things \it must \rm be true: if the emitting region shared the primary's motion and our measurements accurately reflected this motion, our RV measurements would reliably indicate the motion of the primary and no phase shift would be observed.  Any explanation of the phase shift must therefore invoke at least one of the following: a \it physical displacement model \rm, in which the emitting region does not share the motion of the primary, or a \it measurement offset model \rm, in which there is an offset between the measured velocity and the actual velocity of the emitting region.  Many measurement offset models are possible, differing in details such as the nature of the asymmetry producing the offset; we will show, however, that all these models have some predictions in common, which may be tested with existing data.  Similarly, there are a number of testable predictions common to all physical displacement models; it is therefore possible to distinguish between these two explanations, without additional assumptions on the nature of the disk.  In this section, we discuss physical displacement models and give an introduction to measurement offset models, which will be discussed further in subsequent sections.

\subsection{Physical Displacement Models}

We begin by considering physical displacement models, which have been widely adopted in the literature.  In their simplest form (e.g. Stover 1981, Robinson 1992), physical displacement models assume that the line emission is dominated by a small region of the disk, the ``hot spot" whose existence has been well-established by doppler tomography (e.g. Shahbaz et al., 2004) and eclipse maps (e.g. Groot et al, 2001, henceforth G01).  The hot spot trails behind the primary, so the emission line RV curves track the motion of an emission source which is out of phase with the true motion of the primary.  This simple formulation of physical displacement models assumes that the observed radial velocity is equal to the radial velocity of the hot spot.  However, the observed radial velocity of the emitting region will be the radial velocity of the gas within the region, which may be different from the bulk motion of the emitting region.   Indeed, if we assume that disk gas follows Keplerian orbits around the primary, then it passes quickly in and out of the spot, sometimes moving an order of magnitude faster than the spot itself; doppler tomography indicates that, in fact, the gas in the spot region has these high velocities (e.g. Shahbaz et al. 2004).  Physical displacement models assuming a Keplerian disk are therefore contradicted by the fact that the amplitudes of emission line RV curves are far smaller than the Keplerian velocities which would be observed if the hot spot overwhelmed all other line emission.  

A spectrophotometric study of SW Sex was presented by G01, which invokes an alternative to the standard physical displacement model of Stover (1981).  They recognize that, within the framework of a displacement model, their RV curve reflects the motion of gas within the spot rather than the motion of the spot itself.  They conclude that gas in the hot spot region must have the velocity necessary to explain the observed emission line RV curve.  They present an eclipse map which physically locates the hot spot in the disk, and find that the derived velocity of gas in the spot is very different from the Keplerian velocity, in both direction and magnitude.  They conclude that gas in the emitting region is decoupled from the Keplerian flow, and is part of a vertically extended emitting region. G01 do not speculate about how such low-velocity gas can be stable so near to the white dwarf, but their conclusion is unavoidable if one adopts a physical displacement model for the system and recognizes that the measured radial velocity will be the velocity of gas in the spot.

We now consider physical displacement models in the context of observed line profiles.  We begin by noting that physical displacement models cannot apply to systems showing double-peaked emission lines, since it is clear that we are seeing emission from the full disk in these systems.  Though double-peaked systems often have hot spots, the spots clearly do not dominate all other line emission as is hypothesized by displacement models.  Even in single-peaked systems, the large line widths observed in disk emission lines pose a challenge to physical displacement models: without exception, emission lines from accreting binaries have widths consistent with rotational broadening due to disk gas in Keplerian orbits around the primary.  If the hot spot dominates all line emission, we would expect the profile to be dominated by a component whose line width was much smaller than the rotational broadening of the full disk; even if there is a weaker, broad component due to the full disk, physical displacement models require that the spot emission overwhelms the disk component.  The only way to explain the uniformly large line widths would thus be to assume that local broadening (turbulence)  in the spot region is comparable to the Keplerian velocity (Mach $\sim 50$ using values given in G01).  Such extreme turbulence, particularly in combination with the sub-Keplerian disk velocities discussed above, requires a detailed physical explanation in order to be credible.

Finally, we consider eclipse studies of objects showing a phase shift, e.g G01 and Thoroughgood et al. (2005).  The RV curves in these papers show a rotational disturbance, in which RVs measured during the eclipse are first redder than expected (while blueshifted gas is eclipsed), then bluer than expected (while redshifted gas is eclipsed).  The rotational disturbance shows that line emission is coming from a physically extended region, rotating in the same direction as the accretion disk.  This would be hard to explain with physical displacement models, in which emission is assumed to be dominated by a small part of the disk.  In particular, the appearance of a rotational disturbance in a system with single-peaked emission lines (G01) indicates that highly localized emission does not provide a simple explanation for single-peaked lines.  The rotational disturbances in G01 and Thoroughgood et al.  are roughly symmetrical, with the red disturbance having amplitude and duration comparable to the blue one.  If, as displacement models require, the measured RV depended primarily on emission from one part of the disk, then the measured RV would show little effect from the eclipse of other parts of the disk, but would change suddenly when the line-emitting region was eclipsed.  The smoothness and symmetry of the observed rotational disturbance shows that no one region dominates the measured radial velocity, inconsistent with physical displacement models.  We conclude that displacement models are contradicted by a range of observations.
 
\subsection{Measurement Offset Models}

We have argued in the introduction of this section that any explanation of the phase shift must invoke either a physical displacement model or a measurement offset model; since we find the former to be inconsistent with existing data, we now turn our attention to the latter.  Measurement offset models were introduced by Smak (1970, henceforth S70), but have received little attention in the literature since then.  The premise of measurement offset models is that line emission comes from the full disk, centered on the primary, but that small anisotropies in the disk cause the measured radial velocity to differ from the actual velocity of the emitting region.  For example, if orbits of gas in the emitting region are not circular, the asymmetric velocity field will cause the measured radial velocity to differ from the bulk motion of the disk (Paczynski, 1977), producing a measurement offset.  If such an asymmetry is locked to the orbital geometry, for example if it is due to the impacting accretion stream, the measurement offset would change with our viewing angle of the disk and would therefore be a  function of phase.  This phase-dependant measurement offset could potentially alter the amplitude, phasing, and shape of the observed radial velocity curve.  

Large asymmetries are not required to produce a significant measurement offset: the velocity of disk gas is typically large compared to the velocity of the primary, so even a small asymmetry in the disk can produce a large effect on the measured RV of the primary.  For example, if the Keplerian velocity of disk gas is an order of magnitude larger than the velocity of the primary, a 10\% velocity anisotropy (e=0.1 if the disk is elliptical) will produce a measurement offset equal to the true velocity of the primary; smaller asymmetries will still produce considerable effects on the measured RV.  Given the asymmetric lines which are often emitted by accretion disks (e.g. Orosz et al. 1994), there is no reason to assume perfectly symmetrical line-emitting regions.  Moreover, Marsh, Horne \& Shipman (1987) have shown that even a symmetrical profile may not be centered on the velocity of the primary; measurement offsets may thus arise even from lines which give no direct indication of an anisotropic disk.

Hot spots may also produce a measurement offset, and were invoked in the measurement offset model of S70; they are, however, handled differently in these models than in physical displacement models.  Whereas physical displacement models assume that the spot dominates all line emission and thus fully determines the measured velocity, measurement offset models assume that the full disk has strong line emission, and that the spot is a perturbation in a roughly symmetrical emission region.  Thus, measurement offset models invoking a hot spot are not contradicted by the observations discussed in the previous section, e.g. large line widths and the smooth rotational disturbance seen in eclipsing systems.  Unlike G01, measurement offset models also do not require that the gas in the spot is sub-Keplerian: since the spot is treated as a perturbation, the resulting measurement offset will be only a fraction of the velocity of gas in the spot region.  In terms of interpreting emission line RV curves, physical displacement models do not imply any connection between the observed radial velocity and the velocity of the primary, since the gas in the spot can be assumed to have an arbitrary velocity (as in G01). Measurement offset models, by contrast, imply that the measured velocity differs from the true velocity of the primary by a perturbation which might, in principle, be understood and corrected for. 
  
There are many possible physical origins of a measurement offset, especially since a small asymmetry can produce a large effect on the measured velocity.  In addition to the hot spot and eccentric gas orbits discussed above, measurement offsets could be produced by a warped disk, an asymmetrical wind, or an asymmetry in the disk's optical thickness; any asymmetry which is locked to the binary geometry could potentially produce a phase shift.  Ultimately, one might hope to distinguish between the possible physical origins of a velocity offset, but in this paper we assume only that some asymmetry produces a measurement offset, without specifying its nature.  We will show that many aspects of measurement offset models are independent of the physical nature of the underlying asymmetry, and that measurement offset models may therefore be tested and used even if the origin of the measurement offset is unknown.   In the following sections, we will describe the features of measurement offset models which are independent of the origin of the offset, and will test these models against existing data.

\section{Basic Properties of Measurement Offset Models}

In this section, we discuss the fundamental properties common to all measurement offset models, independent of the physical origin of the measurement offset.  We begin by setting the notation which will be used throughout this section.  We assume that the observed radial velocity $v_{obs}$ arises as the sum of the radial velocity of the primary, $v_1$, and a measurement offset, $v_\mathit{off}$.  Since measurement offset models hypothesize that the measurement offset changes with our viewing angle of the accretion disk, we assume that $v_\mathit{off}$, like $v_1$ and $v_{obs}$, is a function of binary phase $\phi$.  We assume that $v_1(\phi)$ is a sinusoid, and write $v_1(\phi)=K_1\sin(\phi)$.  Since we have defined $v_1$ as a radial velocity, $K_1$ differs from the true velocity of the primary by the usual factor of $\sin(i)$.  Authors throughout the literature find that $v_{obs}(\phi)$ is consistent with a sinusoid, and the high time-resolution observations of Thoroughgood et al. (2005) establish that $v_{obs}$ is sinusoidal in at least one system showing a large phase shift.  We henceforth assume that $v_{obs}(\phi)$ is well-approximated by a sine in all cases.  It follows that $v_\mathit{off}(\phi)$ must be sinusoidal: it is the difference between the sinusoids $v_1(\phi)$ and $v_{obs}(\phi)$, and it is a well-known mathematical fact that the sum or difference of two sinusoids of the same period is again a sinusoid.  We therefore write $v_{obs}(\phi)=K_{obs}\sin(\phi - \phi_{obs})$, and $v_\mathit{off}(\phi)=K_\mathit{off}\sin(\phi - \phi_\mathit{off})$.  In our notation, $K_1$ is the quantity needed for a determination of $q$; we seek to understand its relationship to the observables $K_{obs}$ and $\phi_{obs}$.  

 Measurement offset models imply that the observed velocity $v_{obs}(\phi)$ arises as the sum of two sinusoids: the true RV of the primary and a sinusoidally varying measurement offset.  We thus begin our analysis of measurement offset models by discussing the properties of summed sinusoids.  We will see that summed sinusoids produce relationships which would not be expected from the single sinusoid invoked by physical displacement models.  The characteristics of summed sinusoids may therefore be used to test the validity of measurement offset models.   In section 3.1, we give a general discussion of the properties of summed sinusoids and the phase shifts which may come from them.  We will see that that in measurement offset models, small phase shifts do not necessarily correspond to reliable radial velocity measurements.  We will also see that $K_\mathit{off}/K_1$ is a parameter of fundamental importance to measurement offset models.  In section 3.2, we will discuss the physical phenomena which determine this parameter.

\subsection{Properties of a Sinusoidal Offset}
In figure 1, we plot $\phi_{obs}$ as a function of $\phi_\mathit{off}$ for various values of $K_\mathit{off}/K_1$.  The relationship between these three quantities is of fundamental importance to measurement offset models.  Figure 1 shows that for a given value of $K_\mathit{off}/K_1<1$, there is a maximum possible phase shift.  A relatively large range of $\phi_\mathit{off}$ yields observed phase shifts near the maximum possible $\phi_{obs}$, implying that the maximum possible phase shift is also the most likely to be observed. The maximum phase shift increases with $K_\mathit{off}/K_1$, and approaches $\phi_{obs}=0.25$ as $K_\mathit{off}/K_1$ approaches 1.  Additionally, for any given value (or range) of $K_\mathit{off}/K_1<1$ there is a corresponding upper bound on $\phi_{obs}$; thus, though $K_\mathit{off}/K_1$ is not directly observable, its range may be indicated by the range of observed phase shifts.  For $K_\mathit{off}/K_1>1$, any phase shift is possible.    

\begin{figure} \figurenum{1}
\plotone{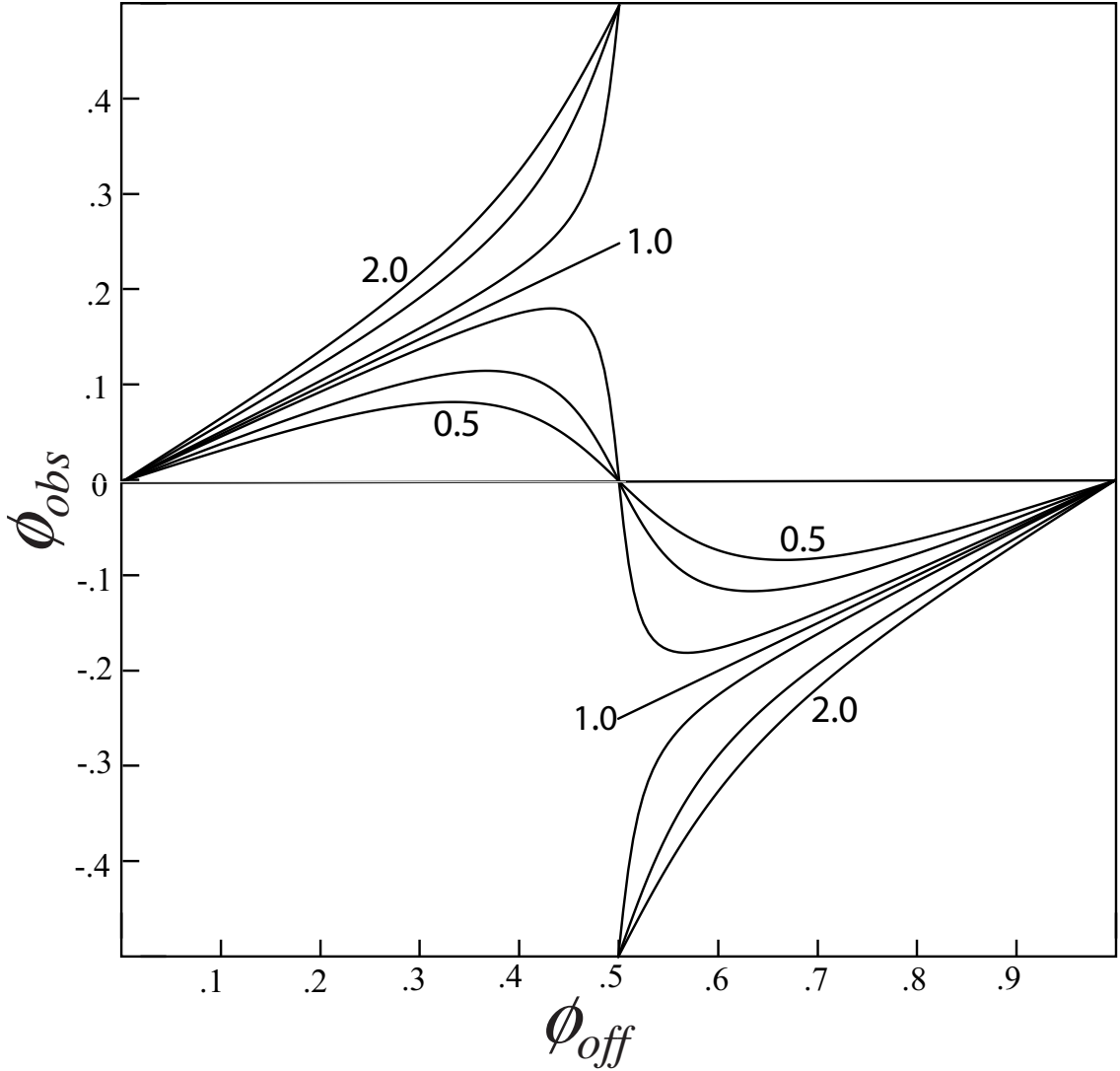}
\caption{Observed phase shifts due to a sinusoidal velocity offset, plotted as a function of $\phi_\mathit{off}$, for various values of $K_\mathit{off}/K_1$; these parameters are defined in section 3.1.  Shown are, from top to bottom on the left hand side and bottom to top on the right hand side, $K_\mathit{off}/K_1=2, 1.5, 1.1, 1, 0.9, 0.66$, and $0.5$.  For each value of $K_\mathit{off}/K_1<1$, there is a corrsponding maximum possible phase shift.  This phase shift is also the most common, with progressively smaller phase shifts coming from progressively smaller ranges of $\phi_\mathit{off}$ for a given value of $K_\mathit{off}/K_1$.  It is also apparent that the most extreme phase shifts, positive and negative, come from similar parameters, namely $\phi_\mathit{off}\approx 0.5$ and $K_\mathit{off}/K_1\gtrsim 1$.  The sign is a function only of whether $\phi_\mathit{off}>0.5$ or $\phi_\mathit{off}<0.5$.  Measurement offset models therefore predict no fundamental difference between objects with positive phase shifts and those with negative phase shifts: a slight change in the underlying parameter $\phi_\mathit{off}$ may change the sign of the observed shift, turning a large, positive shift to a large, negative one.} 
\end{figure}

Figure 1 also shows that the sign of the phase shift depends entirely on the value of $\phi_\mathit{off}$: positive phase shifts occur when $\phi_\mathit{off}<0.5$, and negative phase shifts occur when $\phi_\mathit{off}>0.5$.  Moreover, the transition from large positive phase shifts to large negative ones is quite sudden, becoming discontinuous at $K_\mathit{off}/K_1=1$.  The largest phase shifts, both positive and negative, occur when $K_\mathit{off}/K_1\geq1$ and $\phi_\mathit{off}\approx 0.5$; the sign of the observed shift will be positive if $\phi_\mathit{off}$ is just under 0.5, and negative if it is just over 0.5.  Measurement offset models therefore predict a rather counterintuitive distribution of observed phase shifts: negative phase shifts may be quite large, and the most extreme phase shifts, positive and negative, are likely to be observed in otherwise similar objects. 

Finally, figure 1 shows that for a given value of $K_\mathit{off}/K_1<1$, the smallest phase shifts occur when $\phi_\mathit{off}\approx0$ or $\phi_\mathit{off}\approx0.5$, in other words when $v_1(\phi)$ and $v_\mathit{off}(\phi)$ are nearly in phase or in antiphase.  In these cases, the maxima and minima of the two summed curves line up, and though the offset velocity has little effect on the observed phasing, it may still have a large effect on the observed amplitude.  Thus, small phase shifts are not an indicator of objects for which $K_{obs}$ can reliably be used to determine the mass ratio.  On the contrary, it is possible that the smallest phase shifts may be associated with the objects in which $K_{obs}$ is the least robust as an indicator of $K_1$.   However, small phase shifts may also be the result of small values of $K_\mathit{off}/K_1$, in which case the observed semiamplitude would be robust.  Understanding the parameter $K_\mathit{off}/K_1$ is thus essential to the interpretation of disk emission line RV curves, specifically to understanding when $v_{obs}$ may be trusted as a measure of $v_1$. 

\subsection{The Parameter $K_\mathit{off}/K_1$}
We have shown in the preceding section that the parameter $K_\mathit{off}/K_1$ is of fundamental importance to measurement offset models.  In this section, we develop a framework for for understanding $K_\mathit{off}/K_1$ in terms of physically meaningful quantities.  We will show that $K_\mathit{off}$ depends on two things: (1) the typical radial velocity of line-emitting disk gas relative to the primary ($v_{disk}$), and (2) the dimensionless asymmetries present in the disk.  We begin with an intuitive description of how these two dependencies arise.  (1) As the velocity of disk gas increases, the observed line profile becomes wider, making it harder to pinpoint the primary's true radial velocity.  Indeed, if all disk gas moved slowly relative to the primary, the observed line would be thin and near the true velocity of the primary.  Any measurement offset would then be small, regardless of the asymmetries which might exist in the emitting region.  If the radial velocity of disk gas relative to the primary doubles, the line width---and the potential for error also doubles. (2) In a perfectly symmetrical disk, there would be no offset velocity: a symmetrical disk, centered on the primary, will produce a symmetrical emission line, centered on the radial velocity of the primary.  In this case, the  center of the emission line, easily determined by any means, will be the true radial velocity of the primary.  The existence of a measurement offset thus requires an asymmetry in either the velocity field of the disk (e.g. Paczynski, 1977) or the intensity of the line emission (e.g. Smak, 1970).  The larger this asymmetry, the larger the resulting offset will be.

In section 3.2.1, we show that for disks with identical dimensionless asymmetries, viewed at low inclination,  $K_\mathit{off}$ is proportional to the velocity of disk gas.  We will show that under certain assumptions, the $K_\mathit{off}$-$v_{disk}$ relationship is equivalent to a relationship between $K_\mathit{off}/K_1$ and $q$.  In section 3.2.2, we define the ``asymmetry  parameter" $A$, which characterizes the dimensionless asymmetries in a disk.  $A$ will be defined so that $K_\mathit{off}/K_1$ is completely determined by $A$ and $q$.

\subsubsection{$K_\mathit{off}/K_1$ as a Function of Disk Gas Velocity}

In this subsection, we will study the role of the observed radial velocity of disk gas in determining $K_\mathit{off}/K_1$.  In order to isolate the effect of velocity, we will consider disks with identical dimensionless asymmetries.  When we say that two disks have identical dimensionless asymmetries, we mean two things: (1) their velocity fields are identical, up to a scaling by the maximum radial velocity in each disk, and (2) the intensity of their line emission as a function of position is identical, up to a scaling by the maximum intensity in the given line.  If two disks have the same dimensionless asymmetries, they therefore have identical hotspots and velocity anisotropies, differing only by an overall scaling of either velocities or intensity.  

We now show that for disks with identical dimensionless asymmetries, viewed at low inclination, $K_\mathit{off}$ is proportional to the radial velocity of disk gas, which we denote by $v_{disk}$.  If the inclination is low enough to neglect self-absorption, then disks with identical dimensionless asymmetries will produce emission lines which are identical in shape, differing only by an overall scaling of the height and width of the line; the width scales with $v_{disk}$.  This is equivalent to a relabeling of the axes, with units on the velocity axis scaling with $v_{disk}$.  Since we have taken all velocities to be measured relative to the primary's velocity, $v_1$ is the fixed point of this relabeling of the axes.  $v_1$ will therefore remain fixed relative to the profile shape for any two disks with the same dimensionless asymmetries.  $v_{obs}$ is also fixed relative to the shape of the profile.   Since $v_{obs}$ and $v_{1}$ are both fixed relative to the shape of the profile, their difference scales with the width of the profile, which is proportional to $v_{disk}$.  We therefore have $v_{disk}\propto v_{obs}-v_1$.  Since $v_{obs}-v_1:=v_\mathit{off}$, and $K_\mathit{off}$ is just the amplitude of $v_\mathit{off}(\phi)$, we have that

\begin{equation}
{K_\mathit{off} \propto v_{disk}}.
\end{equation}

To make equation 1 more specific, we need to choose a 
well-defined measure of disk velocity to use as a 
scaling factor.  The outer disk dominates
optical emission from compact binary accretion because
it has the most area and the gas coming from the
companion start is already heated to temperatures
associated with optical blackbody radiation.  Thus we
choose as our disk reference velocity $v_{tide}$, the Keplerian
velocity at the outermost edge of the disk, where
tidal truncation occurs.  The location of the tidal
radius is a function of the mass ratio $q$ for a given binary separation, so it is
natural to express $v_{tide}$, and by extension $K_\mathit{off}$,
in terms of $q$.

We will now show that $K_\mathit{off}/K_1$ is a function of $q$ for disks with identical dimensionless asymmetries in which the velocity of line-emitting disk gas scales with the tidal velocity.  Under these assumptions, we have from (1) that 
\begin{equation}
{K_\mathit{off}\propto v_{disk}\propto K_{tide},}
\end{equation}
where $K_{tide}:=v_{tide}\sin(i)$, and $v_{tide}$ is the usual Keplerian velocity of gas at the tidal radius.   We let $K_{tot}$ denote the sum of the semiamplitudes of the radial velocity curves of the primary and the secondary, so that $K_{tide}$, $K_1$, and $K_{tot}$ all differ from the respective space velocities by the same factor of $\sin(i)$.  We then have from (2), Kepler's laws and the conservation of momentum that 
$$
{K_\mathit{off}\over K_1}\propto{K_{tide}\over K_1}={K_{tide}\over K_{tot}}{K_{tot}\over K_1} = ({M_1/r_{tide} \over (M_1+M_2)/a})^{1\over2}(1+q^{-1})=(1+q)^{-{1\over 2}}({a\over r_{tide}})^{1\over 2}(1+q^{-1}),
$$
where $r_{tide}$ is the disk's tidal radius, $M_1$ and $M_2$ are the masses of the two stars, and $a$ is the semimajor axis of the binary orbit.  Using the fact (Frank, King and Raine (2002), henceforth FKR, p. 141) that the radius $r_1$ of the primary's Roche lobe is given by $r_1\approx 1.1 r_{tide}$, we thus have 
$$
({K_\mathit{off}\over K_1})^2\propto q^{-1}{a\over r_1}(1+q^{-1})
$$
We now use the following approximations, both good for $0.1<q<0.8$ (FKR, p.54):
$$
{a\over r_2}=2.16 (1+q^{-1})^{1\over 3}
$$
$$
{r_2\over r_1}=q^{0.45}.
$$
Combining the above three equations yields
\begin{equation}
{{K_\mathit{off}\over K_1}\propto(1+q^{-1})^{2\over 3}q^{-0.275}.}
\end{equation}

We have show in (1) that, for disks with identical dimensionless asymmetries viewed at low inclination, $K_\mathit{off}$ scales with the velocity of disk gas.  By additionally assuming that disk gas velocities scale with the tidal velocity (2), we derive (3), which expresses $K_\mathit{off}/K_1$ as a function of $q$.  Though there is no explicit inclination dependance in (3), we assumed in our derivation that the inclination was low enough for self-absorption to be negligible.  The $q$-dependance expressed in (3) may therefore break down in high-inclination systems.  For low-$i$ systems, the $q$-dependance of $K_\mathit{off}/K_1$ is derived from and equivalent to the dependance of $K_\mathit{off}$ on disk velocity.  

\subsubsection{$K_\mathit{off}/K_1$ and the Asymmetry Parameter $A$}

In this section, we present a formalism for describing the role of dimensionless asymmetries in determining $K_\mathit{off}/K_1$.  For disks with identical dimensionless asymmetries, we have shown in (3) that $K_\mathit{off}/K_1$ is proportional to a function of $q$.  If we make the proportionality constant explicit, (3) states that for a given set of dimensionless asymmetries, there is some value of $A$ such that all disks with those asymmetries follow the relation 
 
\begin{equation}
{K_\mathit{off}\over K_1}= A(1+q^{-1})^{2\over 3}q^{-0.275}.
\end{equation}

This equation defines the asymmetry parameter $A$.  For a given set of dimensionless asymmetries, $A$ is constant.  Its value depends only on the dimensionless asymmetries present in a disk.  Unlike standard measures of dimensionless asymmetry (e.g. eccentricity), $A$ includes all types of dimensionless asymmetry, quantifying the total asymmetry in a disk in terms of its impact on observed radial velocities.  A given $A$-value may therefore be caused by a range of different physical anisotropies.  We will not consider the possible physical origins of $A$ in this paper, using $A$ only as a tool to understand $K_\mathit{off}/K_1$ and test measurement offset models.

We have shown that $K_\mathit{off}/K_1$ varies as a function of the mass ratio $q$ and the geometry of the underlying disk, whose dimensionless asymmetry is characterized by the asymmetry parameter $A$.  There may be an additional dependence on $i$; the inclination effect, if any, should be relevant only at inclinations large enough for self-absorption to be significant.  As discussed in section 3.1, figure 1 shows that there is a theoretical upper bound on observed phase shifts, as a function of $K_\mathit{off}/K_1$.  This, combined with the dependence of $K_\mathit{off}/K_1$ on $q$, will allow us to test measurement offset models against existing data.  

\section{Variations in $K_\mathit{off}/K_1$: Empirical Results}
In this section, we use data from the literature, together with the results of the previous section, to test measurement offset models and to study the parameter $A$.  In section 4.1, we present a comprehensive collection of existing phase shift data from the literature.  This gives us sufficient data to separate the effects of $q$ and $i$ on the phase shift, and to detect the presence of any upper bound on $\phi_{obs}$ as a function of both these quantities. We begin in section 4.2 by determining $A$ directly in 12 objects where this can be done; the work of this section suggests several patterns in the value of $A$, most notably that it is constant when $i$ is small and $q$ is not near unity.  In 4.3, we interpret our results on $A$ in terms of an upper bound on observed phase shifts, and test this bound using our full sample of phase shifted emission line radial velocity curves.  We find that the full sample is in agreement with the conclusions drawn in section 4.2.  The agreement between these two sections supports measurement offset models, as the methods of the two sections are logically independent and are connected only by the adoption of these models.  In section 4.4, we discuss the physical meaning of our results, and their implications for models of the phase shift in various objects.  

\subsection{Collection of Phase Shift Data}
In order to study the dependence of observed phase shifts on $q$ and $i$, we have compiled a comprehensive listing of accreting binaries with measured Balmer line phase shifts, shown in table 1.  Also included in this table are the mass ratio and, when available, the mass of the secondary star, the period of the system, the inclination, and the type of object.  Table 1 includes all accreting binaries with published disk Balmer line RV curves and an alternate determination of the ephemeris, even in cases where no phase shift is observed; it is thus also a complete listing of accreting binaries which are known to have zero phase shift to within observational errors.

Measuring a phase shift requires both an emission line radial velocity curve and an independent determination of the time of conjunction of the two stars.  For the former, we have included only measurements made with Balmer lines, using H$\alpha$ when available; for the latter, we have included results using either absorption line radial velocity curves or eclipses, as these are generally in agreement when both are available.  We have also included determinations of phase shifts in which the measurement of the time of conjunction and the emission line radial velocity curve appear in different papers.  Finally, several of the listed phase shifts were obtained from papers in which an ephemeris was determined and a sinusoidal fit to emission line curves was shown graphically, but the parameters of the fit were not explicitly given.  In these cases, we measured the phase shift in a graphics program, but no error is listed in table 1, as the uncertainty in the fit itself is uncertain.  

The objects listed in table 1 were found by doing a literature search on every object with a measured mass ratio $q$ listed in the RK Catalogs of CVs and LMXBs (Ritter \& Kolb, 2003, henceforth RKcat).  Not all objects with a measured mass ratio have the necessary data to determine a phase shift, but every object for which both emission line and absorption line radial velocity curves exist has some measurement of the mass ratio.  We therefore believe that our sample is nearly complete, possibly missing only those objects in which eclipse observations and emission line radial velocity curves exist, but no absorption line study or other determination of the mass ratio has been done.  In many cases, the data needed to determine a phase shift exist, but more robust measurements of $q$ have been made and are referenced in RKcat; in cases where more than one determination of $q$ exists, we list the value given in RKcat.  Table 1 will be used in the following sections to test the predictions made by measurement offset models.

\begin{deluxetable}{lllllllll}
\tablecaption{Accreting Binaries with Phase Shift Data\tablenotemark{a}}
\tabletypesize{\footnotesize}
\tablehead{
\colhead{name} & \colhead{$\phi_\mathit{off}$} & \colhead{$\sigma \phi_\mathit{off}$} & \colhead{$q={M_2\over M_1}$} & \colhead{$M_2$(M$_\odot$)} & \colhead{$P_{orb}$(d)} & \colhead{$i$($^\circ$)} & \colhead{type\tablenotemark{b}} & \colhead{ref.\tablenotemark{c}}}
\startdata
GK Per & 0 & \nodata & $0.47\pm0.18$ & \nodata & 1.996803 & \nodata & DN & 1\\

TT Crt & 0 & \nodata & 0.59 & \nodata & 0.268352 & \nodata & DN & 2($\phi_{obs}$, $q$)\\ 

RY Ser & 0 & \nodata & $0.83\pm0.14$ & \nodata & 0.3009 & \nodata & DN & 2\\

J2044-0459 & 0 & \nodata & $0.36\pm0.09$ & \nodata & 1.68 & \nodata & \nodata & 3\\

CH UMa & 0 & \nodata & $0.43\pm0.07$ & $0.96\pm0.01$ & 0.343184 & $21\pm4$ & UG & 2\\

IX Vel & 0.01 & 0.01 & $0.65\pm0.04$ & $0.53\pm.09$ & 0.193927 & $60\pm5$ & UX & 5\\

AH Her & 0.01 & 0.01 & $0.80\pm0.05$ & $0.76\pm.08$ & 0.258116 & $46\pm3$ & ZC & 6\\

GY Cnc & 0.013 & 0.008 & $0.38\pm0.03$ & $0.38\pm0.06$ & 0.175442 & $77.3\pm0.9$ & DN & 7\\

U Gem & 0.015 & 0.011 & $0.362\pm0.01$ & $0.39\pm0.02$ & $0.176906$ & $69\pm2$  & UG & 9\\

BV Cen & 0.016 & 0.006 & $1.09\pm0.07$ & $0.9\pm0.1$ & 0.610108 & $62\pm5$ & UG  & 10\\

UY Pup & 0.02 & \nodata& $0.91\pm0.08$ & \nodata & 0.479269 & \nodata & ZC & 2\\

V392 Hya & 0.025 & \nodata & $0.56\pm.05$ & \nodata & 0.324952 & \nodata & UG? & 3 \\

J0813+4528 & .026 & \nodata & $0.59\pm0.05$& \nodata & 0.2890 & \nodata & UG? & 2\\

V603 Aql & 0.03 & 0.01 & $0.24\pm0.05$ & $0.29\pm0.04$ & 0.1385 & $13\pm2$ & SH & 11\\

CN Ori & 0.03 & 0.01 & $0.66\pm0.04$ & $0.49\pm0.08$ & 0.163199 & $67\pm3$ & UG & 9\\

J1951+3716 & 0.04 & \nodata & $0.67\pm0.09$ & \nodata & 0.492 & \nodata & NL? & 3\\

VY Scl & 0.04 & 0.03 & $0.28\pm0.09$ & $0.43\pm.13$ & $0.2323$ & $30\pm10$ & VY & 12\\

AY Psc & 0.05 & 0.03 & 0.45 & 0.59 & 0.217321 & 74 & ZC & 13($q, \phi_{obs}, M_2, i$)\\

HR Del & 0.05 & \nodata & $0.83\pm0.07$ & $0.55\pm0.03$ & 0.214165 & $40\pm2$ & Nb & 14\\

EI Psc & 0.05 & 0.02 & $0.19\pm0.02$ & $0.13\pm0.03$ & 0.044567 & $50\pm5$ & SU & 15\\

EM Cyg & 0.057 & .011 & $0.88\pm0.05$ & $0.99\pm0.12$ & 0.290909 & $67\pm2$ & ZC & 9\\

V347 Pup & 0.06 & \nodata & $0.83\pm0.05$ & $0.52\pm0.06$ & 0.231936 & $84\pm2.3$ & NL & 16\\

V630 Cas	& 0.06 & 0.013 & $0.18\pm0.02$ & $0.18\pm0.02$ & 2.56387 & $72.5\pm5.5$ & DN & 17\\

DX And & 0.066 & 0.017 & $0.67\pm0.09$ & \nodata & 0.440502 & $45\pm12$ & UG & 18\\

IP Peg & 0.067 & 0.008 & $0.45\pm0.04$ & $0.42\pm0.08$ & 0.158206 & $81.8\pm0.9$ & UG & 9\\

AC Cnc & 0.07 & \nodata & $1.02\pm.04$ & $0.77\pm0.05$ & 0.300478 & $75.6\pm.7$ & SW & 19\\

V363 Aur & 0.075 & \nodata & $1.18\pm.07$ & $1.06\pm0.11$ & 0.32124 & $69.7\pm.4$ & SW & 19\\

V893 Sco & 0.08 & 0.012 & 0.19 & 0.175 & 0.075962 & 72.5 & SU? & 20\\

AT Ara & 0.084 & 0.01 & $0.79\pm0.04$ & $0.42\pm0.1$ & 0.3755 & $38\pm5$ & UG & 21\\

V426 Oph & 0.09 & 0.02 & $0.78\pm0.06$ & $0.7\pm0.14$ & 0.2853 & $59\pm6$ & ZC & 22\\

Z Cha & 0.072 & 0.019 & $0.15\pm0.004$ & $0.125\pm0.014$ & 0.074499 & $81.8\pm0.1$ & SU & 23\\

HT Cas & 0.094 & 0.012 & $0.15\pm0.03$ & $0.09\pm0.02$& 0.073647 &$81\pm 1$& SU & 24\\

XN Mus 91 & 0.10 & .016 & $0.13\pm0.02$ & $0.85\pm0.25$& 0.432602 & $54\pm1.5$& LMXB & 25\\

DV UMa & 0.10 & 0.02 & $0.151\pm0.001$ & $0.157\pm.004$ & 0.085853 & $84.2\pm0.1$ & SU & 9\\

SW Sex & 0.10	 & \nodata & $0.6\pm0.2$ & $0.33\pm.06$ & 0.134938 & $82\pm7$ & SW & 26($q, \phi_{obs}, M_2, i$)\\

OY Car & 0.117 & 0.006 & $0.102\pm0.003$ & $0.07\pm0.002$ & 0.063121 & $83.3\pm0.2$ & SU & 9\\

IY UMa & 0.12 & 0.03 & $0.125\pm0.008$ & $0.1\pm0.01$  & 0.073909 & $86\pm1$ & SU & 27\\

A0620-00 & 0.12 & 0.013 & $0.062\pm0.024$ & $0.68\pm0.18$ & 0.323016 & $41\pm3$ & LMXB & 24\\

V838 Her & -0.13 & 0.01 & $0.85\pm0.13$ & $0.074\pm0.01$ & 0.297635 & $84\pm6$ & Na & 28\\

WX Cet  & 0.130 & 0.007 & 0.084 & $0.047\pm0.013$ & 0.05829 & \nodata & SU & 29\\

V2051 Oph & 0.132	 & 0.02 & $0.19\pm0.02$ & $0.15\pm0.03$ & 0.062428 & $83\pm2$ & SU & 9\\

WZ Sge  & 0.133  & 0.006  & $0.06\pm0.01$  & $0.045\pm.003$  & 0.056688  & $75.9\pm0.3$  & SU  & 30($q$), 9($\phi_{obs}$)\\

OY Ara & 0.138 & 0.011 & $0.41\pm0.06$ & $0.34\pm0.01$ & 0.155466 & $74.3\pm1.2$ & Na & 31($q, \phi_{obs}$)\\

V1315 Aql & 0.14 & 0.03 & $0.34\pm0.13$ & $0.30\pm0.01$ & 0.139690 & $82\pm4$  & SW & 32\\

BH Lyn & 0.15 & \nodata & $0.45\pm0.11$ & $0.33\pm0.14$ & 0.155875 & $79\pm4$ & SW & 33\\

XN Vel 93 & 0.16 & \nodata & $0.14\pm0.02$ & 0.6 & 0.285206 & 78 & LMXB & 34\\

BT Mon & 0.17 & 0.01 & $0.84\pm0.06$ & $0.87\pm0.045$ & 0.324952 & $82\pm3$ & SW? & 35\\

GRO J0422+32 & 0.20 & \nodata & $0.11\pm0.03$ & $0.46\pm0.31$ & 0.212160 & $45\pm2$ & LMXB & 36\\

GS 2000+25 & 0.22 & \nodata & $0.041\pm0.008$ & $0.33\pm0.08$ & 0.344087 & $64\pm1.3$& LMXB & 37\\

V2107 Oph  & -0.24 & \nodata & $<0.053$ & \nodata & 0.521 &  \nodata & LMXB & 38($q$), 39($\phi_{obs}$)\\

EX Dra & 0.27 & 0.03 & $0.75\pm0.01$ & $0.56\pm0.02$ & 0.209937 & $84.2\pm0.6$ & UG & 40\\

\enddata
\tablenotetext{a}{Included are Cataclysmic Variables and Low-Mass X-ray Binaries for which there are published Balmer line radial velocity curves and either absorption line radial velocity curves or eclipse observations, allowing the detection of a phase shift $\phi_\mathit{off}$.  There is no cut based on inclination, and all objects are included even if the data show $\phi_\mathit{off}$ consistent with zero.  $\sigma \phi_\mathit{off}$ denotes the error on $\phi_\mathit{off}$.  See text for details.} 

\tablenotetext{b}{We denote Low Mass X-ray Binaries by LMXB, and follow the CV classifications given in RKCat (RItter \& Kolb, 2003): DN=Dwarf Nova; Na=Fast Nova with no sublassification; Nb=Slow Nova with no subclassification; Nr=Recurrent Nova; NL=Novalike Variable, no subclassification; SH=non-SU Uma star showing superhumps; SU=SU UMa star; SW=SW Sex star; UG=U Gem or SS Cyg star; UX=UX UMa star; VY=VY Scl Star; ZC=Z Cam star}

\tablenotetext{c}{Unless otherwise noted, references given are for $\phi_\mathit{off}$ only, and all other values are from Ritter \& Kolb, 2003.}

\tablerefs{1. Crampton, Cowley, \& Fisher 1986, 2. Thorstensen, Fenton, \& Taylor 2004, 3. Peters \& Thorstensen 2005, 4. Beuermann, Stasiewski, \& Schwope 1992, 5. Beuermann \& Thomas 1990, 6. Horne, Wade, \& Szkody 1986, 7. Thorstensen 2000, 9. Mason et al. 2000, 10. Gilliland 1982, 11. Arenas et al. 2000, 12. Martinez-Pais et al. 2000, 13. Szkody \& Howell 1993, 14. Kuerster \& Barwig 1988, 15. Thorstensen et al. 2002, 16. Thoroughgood et al. 2005, 17. Orosz, Thorstensen, \& Honeycutt 2001, 18. Bruch et al. 1997, 19. Thoroughgood et al. 2004, 20. Mason et al. 2001, 21. Bruch 2003, 22. Hessman 1988, 23. Marsh, Horne, \& Shipman 1987, 24. Young, Schneider, \& Schechtman 1981, 25. Orosz et al. 1994, 26. Dhillon, Marsh, \& Jones 1997, 27. Wu, Gao, \& Leung 2001, 28. Szkody \& Ingram 1994, 29. Rogoziecki \& Schwarzenberg-Czerny, 2003, 30. Skidmore et al. 2002, 31. Zhao \& McClintock 1997, 32. Dhillon, Marsh, \& Jones 1991, 33. Hoard \& Szkody 1997, 34. Filippenko et al. 1999,  35. Smith, Dhillon, \& Marsh 1998, 36. Filippenko, Matheson, \& Ho 1995, 37. Filippenko, Matheson, \& Barth 1995 38. Harlaftis et al. 1997, 39. Filippenko et al. 1997, 40. Fielder, Barwig, \& Mantel 1997}
\end{deluxetable}
	
\subsection{Direct Determination of $A$}
For those objects in table 1 which have published mass ratios both from emission line radial velocity curves and from some independent technique, it is possible to determine the asymmetry parameter $A$ directly; we begin our study of $A$ with these objects.  Given $q$ and the RV curve of the secondary, the RV curve of the primary, i.e. $v_p(\phi)$, is completely determined: it is in antiphase with the RV curve of the secondary and its amplitude is $qK_2$.  Thus, given $v_{obs}(\phi)$, determined from an emission line study, $v_\mathit{off}(\phi):=v_{obs}(\phi)-v_p(\phi)$ may be computed  directly from its definition.  

We list in table 2 those objects for which an independent determination of the mass ratio exists.  For each object, we list $i$, $q$, $K_{obs}/K_2$ (the mass ratio determined directly from disk emission lines), and $\phi_{obs}$.  We then give $K_\mathit{off}/K_1$, determined using the equation
$$
{K_\mathit{off}\over K_1}\sin(\phi-\phi_\mathit{off})={v_\mathit{off}(\phi)\over K_1}={K_2\over K_1}{v_{obs}(\phi)\over K_2}-{v_1(\phi)\over K_1}=q^{-1}{K_{obs}\over K_2}\sin(\phi-\phi_{obs})-\sin(\phi).
$$ 
 Finally, we give $A$, determined from equation (3) using $K_\mathit{off}/K_1$ and $q$.  The objects in table 2 are listed in order of increasing $A$.

\begin{deluxetable}{lllllllll}
\tablecaption{The Asymmetry Parameter $A$}
\tabletypesize{\footnotesize}
\tablehead{
\colhead{name} & \colhead{$\phi_{obs}$} & \colhead{{$K_{obs}\over K_2$}} & \colhead{$q={K_1\over K_2}$}  & \colhead{${K_\mathit{off}\over K_1}$} & \colhead{$A$} & \colhead{$i$} & \colhead{ref.\tablenotemark{a}}}
\startdata
A0620-00 & $0.12\pm0.013$ & $0.074\pm0.006$  & $0.062\pm0.024$ & $0.83\pm0.25$  & $0.058\pm0.03$ & $41\pm3$ & 1\\
EI Psc & $0.05\pm0.02$ & $0.21\pm0.02$  & $0.19\pm0.02$ & $0.34\pm0.1$  & $0.06\pm0.02$ & $50\pm5$ & 2\\
GS 2000+25 & $0.22$ & $0.05\pm0.025$  & $0.042\pm0.012$ & $1.4\pm0.45$  & $0.068\pm0.027$ & $64\pm1.3$ & 3\\
XN Mus 91 & $0.10\pm 0.016$ & $0.13\pm0.02$  & $0.13\pm0.04$ & $0.62\pm0.16$  & $0.084\pm0.033$ & $54\pm1.5$ & 4\\
WZ Sge & $0.133\pm0.006$ & $0.137\pm0.008$ & $0.06\pm0.01$ & $1.8\pm0.35$  & $0.12\pm0.03$ & $75.9\pm0.3$ & 5\\
GRO J0422+32 & $0.2$ & $0.109\pm0.01$  & $0.116\pm0.075$ & $1.14\pm0.32$  & $0.14\pm0.08$ & $45\pm2$ & 6\\
HT Cas & $0.094\pm0.012$ & $0.26\pm0.02$  & $0.15\pm0.03$ & $1.05\pm0.32$  & $0.16\pm0.06$ & $81\pm1$ & 7\\
IY Uma & $0.12\pm0.03$ & $0.21\pm0.05$  & $0.125\pm0.008$ & $1.2\pm0.35$  & $0.16\pm0.032$ & $86\pm1$ & 8\\
Z Cha\tablenotemark{b} & $0.072\pm0.019$ & \nodata  & $0.149\pm0.004$ & $1.11\pm0.18$  & $0.17\pm0.03$ & $81.8\pm0.1$ & 9\\
U Gem & $0.015\pm0.011$ & $0.6\pm0.2$  & $0.362\pm0.01$ & $0.67\pm0.5$  & $0.21\pm0.08$ & $69\pm2$ &10\\
EM Cyg & $0.057\pm0.011$ & $1.36\pm0.12$  & $0.88\pm0.05$ & $0.7\pm0.14$  & $0.41\pm0.08$ & $67\pm2$ & 11\\
BT Mon & $0.17\pm0.01$ & $2.25\pm0.25$ & $0.84\pm0.11$ & $2.37\pm0.44$  & $1.3\pm0.26$ & $82\pm3$ & 12\\
\enddata
\tablenotetext{a}{The references given are for $q$ only; references for  $K_{obs}\over K_2$ are the same as the references for $\phi_\mathit{off}$ given in table 1, and $i$ is taken from RKcat as in table 1.} 
\tablenotetext{b}{$K_{obs}/K2$ is not given in Marsh, Horne, \& Shipman (1987), who instead give $K_{obs}/K_1$, computed assuming the mass ratio quoted here.  We use this to compute ${K_\mathit{off}\over K_1}$ and $A$, in the same manner as all other objects; see text for details.}

\tablerefs{1. Gelino, Harrison, \& Orosz 2001, 2. Skillman et al. 2002, 3. Harlaftis, Horne, \& Filippenko 1996, 4. Casares et al. 1997, 5. Skidmore et al. 2002, 6. Harlaftis et al. 1999, 7. Horne, Wood, \& Steining, 1991, 8. Steeghs et al. 2003, 9. Wood et al. 1986, 10. Long \& Gilliland 1999, 11. North et al. 2000, 12. Smith, Dhillon, \& Marsh, 1998.}
\end{deluxetable}
 
Table 2 suggests two effects, one which produces much larger values of $A$ in systems with mass ratios near unity, and one which increases $A$ in systems with high inclination.  There is a group of objects with $0.05<A<0.09$ and another group with $0.14<A<0.21$; values of $A$ in the lower range are generally at lower inclinations, and there seems to be a cutoff around $i\sim65-70^\circ$.  WZ Sge ($A=0.12\pm0.03$) is the only object between these groups, and could be consistent with either, but its inclination places it with the high-asymmetry, high-inclination group.  There are also two objects with much larger $A$: EM Cyg ($A=0.41$) and BT Mon ($A=1.3$).  These two objects also have by far the largest mass ratios in table 2, $q=0.84$ and $0.88$.  We speculate that accretion in these objects is not driven by Roche lobe overflow, and that their disks thus have fundamentally different asymmetries, leading to a different value of the asymmetry parameter.  Finally, we note that BT Mon, which has by far the largest $A$ value of any object in table 2, is also the only one which has both large $q$ and high $i$, suggesting that the two effects may be cumulative.  

\begin{figure} \figurenum{2}
\plotone{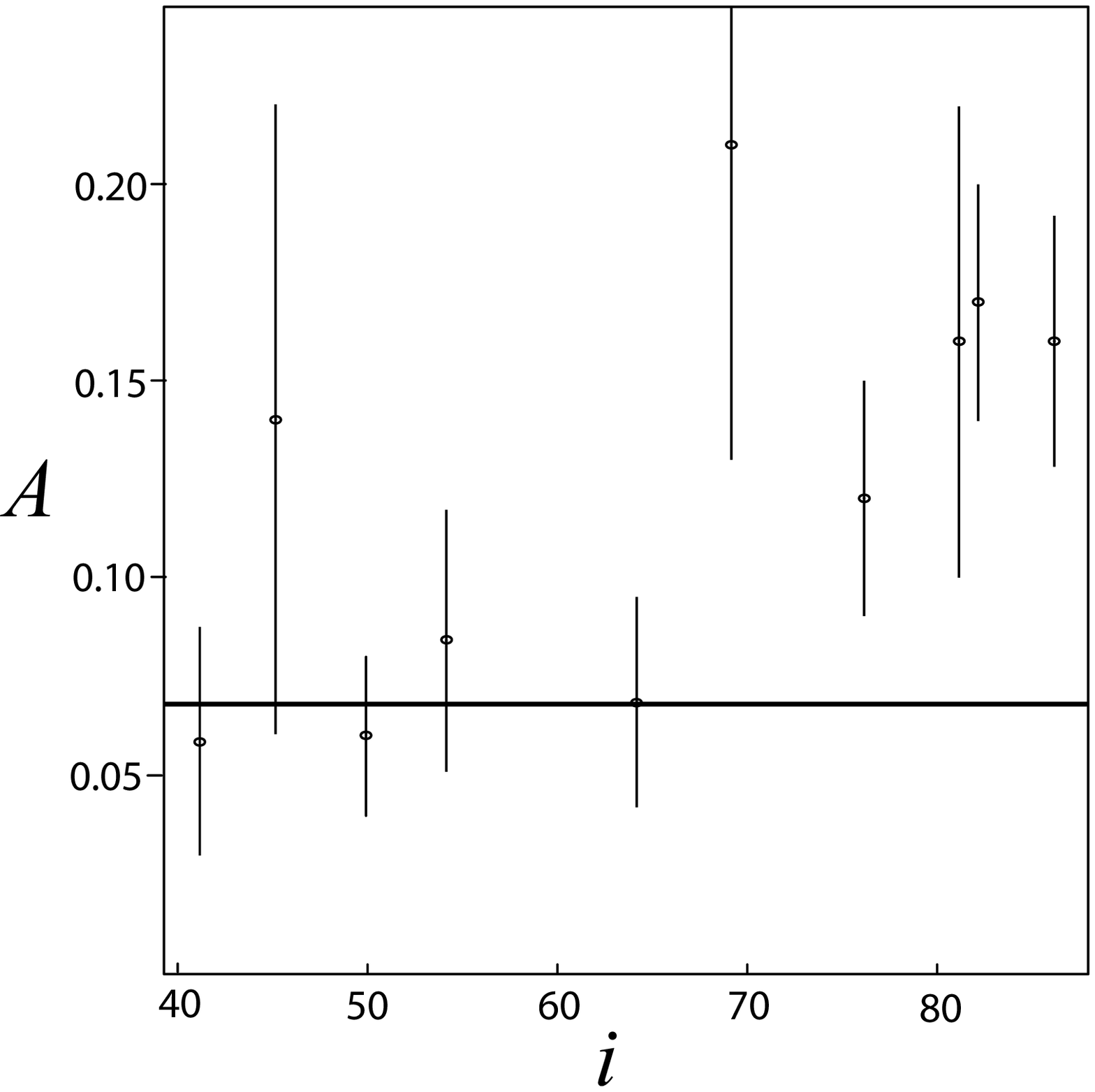}
\caption{The asymmetry parameter $A$ plotted as a function of inclination.  Objects with $q$ near unity show much larger values of $A$ than their low-$q$ counterparts, so they are excluded from this plot in order to isolate the effect of $i$; this excludes only BT Mon and EM Cyg, with all others in table 2 shown.  For each object, $A$ was determined directly by subtracting the predicted motion of the primary from the observed emission line RV curve, then using equation (3) to determine $A$.  Though there is a jump in $A$ at $64^\circ <i<69^\circ$, the data are consistent with being constant at low $i$, as expected if the inclination effect is due to self-absorption.  All values of $A$ in the low-$i$, low-$q$ regime are consistent with $A=0.068$, shown by the horizontal line.  See text for details.} 
\end{figure}

If the inclination dependence of $A$ is due to self-absorption, then we expect the effect to be present only in high-inclination systems, and do not expect any inclination dependence below some critical $i$.  In order to check this, we plot $A$ against $i$ in figure 2, excluding the two objects with high mass ratios, which we believe are part of a separate distribution in $A$.  We consider two models: (1) a simple linear fit, which suggests a continuous change of $A$ with $i$, and (2) a fit by two horizontal lines with a discontinuous break between them, which suggests a discontinuous transition at some critical $i$.  The best linear fit (model 1) has a slope of 0.0024 deg$^{-1}$, an A-intercept of -0.045, and a reduced $\chi^2$ of 0.73.  Fitting the data by two horizontal lines (model 2), we find $A=0.068$ at low inclinations, $A=0.157$ at high inclinations, and a cutoff in inclination constrained to lie between $64^\circ$ and $69^\circ$.  This fit has a reduced $\chi^2$ of 0.52.  The fact that both models give reduced  $\chi^2 <1$ indicates that either the errors in table 2 are systematically overestimated, or the errors on $A$ are non-Gaussian, or both.  However, taking the fits at face value suggests that the discontinuous model, which is physically motivated, is also a slightly better fit.  We note in particular that the objects below $69^\circ$ are strongly clustered in $A$, with a reduced $\chi^2$ of 0.36 around the value of $A=0.068$ used in the fit above, suggesting that $A$ may be constant below $i=68^\circ$.  

There are not enough objects in table 2 to draw any firm conclusions, but our analysis suggests a number of things about the dependancy of $A$ on $q$ and $i$: (1) The data are consistent with a discontinuous change in $A$ as a function of inclination, as would be expected if self-absorption has an effect on $A$ above some critical $i$.  (2) The critical $i$ at which $A$ increases seems to lie between 65$^\circ$ and 69$^\circ$.  (3) The data suggest that $A$ increases sharply in objects with $q\approx 1$, though this is based on only two objects. (4) The effects of $i$ and $q$ may be cumulative: though this conclusion is based on only a single object, it is intriguing that the only high-$i$, high-$q$ object in table 2 also has $A$ a factor of three larger than any other object.  All of these conclusions are highly tentative due to the small numbers of objects under consideration, but they will be refined and further tested in the next section, using a different technique on the full data set of table 1.

\subsection{A Predicted Upper Bound: Confrontation with Observation}

If, as suggested in the previous section, $A$ is constant for low-$i$, low-$q$ systems, this completely determines an upper bound on observed phase shifts in these systems, as a function of $q$: given $A$, equation (3) determines $K_\mathit{off}/K_1$ as a function of $q$, and $K_\mathit{off}/K_1$ determines an upper bound on possible phase shifts, as explained in section 3.1.  We may therefore test the constancy of $A$ and the fundamental soundness of measurement offset models, by comparing this predicted upper bound to the phase shift data in table 1.  If measurement offset models and the results of the previous section are correct, then all objects in the low-$q$, low-$i$ regime should be consistent with the predicted upper bound, but the bound will likely not hold not hold outside this regime.  

\begin{figure} \figurenum{3}
\plotone{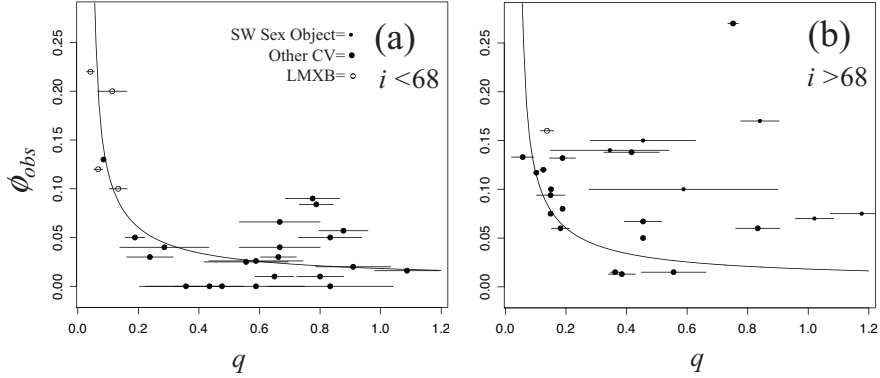}
\caption{Observed phase shift plotted as a function of mass ratio, for (a) systems with $i<68^\circ$ and (b) those with $i>68^\circ$.  Plotted are all objects with nonnegative phase shifts listed in table 1.  The line drawn in both plots is the upper bound on $\phi_{obs}$ derived by adopting measurement offset models and assuming that $A=0.068$ is constant, see section 4.3 for details.  Figure 3a shows that low-$i$, low-$q$ systems are consistent with the upper bound given by $A=0.068$, supporting the constancy of $A$ in this regime.  It also shows the increase in $A$ at $q>0.65$; figure 3b shows that $A$ is generally much larger in high-$i$ systems, probably due to an effect of self-absorption on the observed profile.  All the effects on $A$ which are present in these figures are consistent with those apparent in table 2; the agreement between these two logically independent methods of deriving $A$ supports measurement offset models, as discussed in section 4.4.} 
\end{figure}

In figure 3, we plot $\phi_{obs}$ as a function of $q$ for all objects in table 1, excluding only V2107 Oph and V838 Her, whose large, negative phase shifts will be discussed in section 5 and are clearly not above any positive upper bound.  Objects with $i<68^\circ$, are plotted in figure 3a, and those with $i>68^\circ$ are plotted in figure 3b.  Figure 3a includes objects with no measured inclination; since higher inclination objects are likely to show eclipses and thus have at least a bound on their inclination, we assume that anything with no measurement of $i$ is low-inclination.  Also shown in both figures is the upper bound on the phase shift, derived from equation (3) by assuming $A=0.068$ at all $q$.  This upper bound depends on both the value $A$ and the validity of measurement offset models themselves.  Specifically, the overall shape of the curve--nearly flat at large $q$ and rising sharply to an asymptote at some $q_{asymp}>0$--is fundamental to measurement offset models invoking constant $A$;  the value of $q_{asymp}$  and the height of the upper bound at large $q$ are functions of $A$.  

Figure 3a supports the robustness of measurement offset models and the constancy of $A=0.068$ at  $q<0.65$ and $i<68^\circ$; it tests these hypotheses on the full population of objects with existing phase shift data.  Accreting binaries of all types in the low-$q$, low-$i$ regime are consistent with the derived upper bound.  Deviations from $A\lesssim0.068$ in the low-$q$, low-$i$ regime would be readily apparent in figure 3a.  Indeed, figure 3b shows that the larger values of $A$ typical of high-$i$ objects produce a distribution of phase shifts which is almost entirely above the upper bound for $A=0.068$.   We also note that most observed phase shifts appear to be near the predicted bound.  This is a consequence of the fact that, for a given $K_\mathit{off}/K_1<1$, the maximum possible phase shift is also the most likely to be observed, as discussed in section 3.1. The $q$-value of the vertical asymptote, and the height of the upper bound to the right of the asymptote, are both functions of $A$.  The fact that they are consistent, both with each other and with the value of $A$ derived directly in the previous section, supports the constancy of $A$ throughout the low-$q$, low-$i$ regime.  

We note that the $q-\phi_{obs}$ relationship discussed here explains the $p-\phi_{obs}$ relationship among low-$i$ CVs mentioned by Mason et al. (2000).  This relationship previously had no quantitative explanation.  For CVs, the relatively similar masses of the primaries and the assumption of a main-sequence secondary, produce a correlation between $p$ and $q$ among systems with main sequence secondaries.  Thus, if one considers only CVs, the $q-\phi_{obs}$ relationship yields the $p-\phi_{obs}$ relationship of Mason et al. 

Figure 3 also supports the conclusion that values of $A$ are typically larger in objects with a high inclination or mass ratio.  Any object lying above the line shown in the two figures must have an asymmetry parameter $A>0.068$.  Objects below this line could still have $A>0.068$, since measurement offset models predict only an \it upper\rm \ bound.  Indeed, one of the objects below the bound in figure 3b is U Gem, whose asymmetry parameter, listed in table 2, was computed directly to be $A=0.21\pm0.08$.   Based on this, over half of the high-$q$ and over 80\% of the high-$i$ objects \it must\rm \ have $A>0.068$; the remainder may or may not have large $A$.  There is no clear evidence for (or against) a constant $A$ in the high-$q$ or high-$i$ regimes, either in table 2 or figure 3.  

Finally, we note that figure 3 supports the suggestion from table 2 that the most extreme values of $A$ appear when both $q$ and $i$ are large.  The high-$i$, high-$q$ object EX Dra shows the largest phase shift of any object in figure 3, which suggests a large $A$.  Indeed, its phase shift of 0.27 requires $K_\mathit{off}/K_1>1$ which, at $q=0.75$, implies by equation (3) that $A>0.53$.  The one object in table 2 with a larger value of $A$ is BT Mon, the only high-$i$, high-$q$ object in that table.  Thus, the two largest values of $A$ we know of appear in the two high-$i$, high-$q$ objects in which we have a constraint on $A$, though the constraints on $A$ come from different techniques in the two objects.

\subsection{Discussion}
In sections 4.2 and 4.3, we have studied $A$ via two different methods: direct computation for a few objects in 4.2 and via an upper bound which depends on $A$ in section 4.3.  Both of these techniques are intimately tied to the adoption of measurement offset models: (1) section 4.2 involves a direct computation of $v_\mathit{off}=v_{obs}-v_1$, which is reasonable only if the measurement offset is a meaningful quantity, while (2) the bound of section 4.3 is derived using the properties of summed sinusoids, which are a fundamental feature of measurement offset models.  In addition, measurement offset models predict that the two methods should give consistent results on the properties of $A$, but such consistency would be difficult to explain in the context of other models.  Comparing the results from sections 4.2 and 4.3 therefore provides a test of the robustness of measurement offset models.

The upper bound on $\phi_{obs}$ as a function of $q$, for systems with $q<0.65$ and $i<68^\circ$ provides strong support for measurement offset models.  This bound depends on nothing but the asymmetry parameter $A$ and the adoption of measurement offset models.  The work of section 4.2 suggests that low-$q$, low-$i$ objects have a constant value of $A\approx 0.068$.  In section 4.3, the same value of $A$ produces an upper bound which is consistent with the larger sample of table 1, in the same domain of $q$ and $i$.  The consistency between the two sections is manifest in the fact that, given the work of section 4.2, the bound of figure 3a is derived with no free parameters and no new assumptions.

Sections 4.1 and 4.2 also agree on the range of $i$ and $q$ over which $A$ remains constant.  If we assume that $A$ changes at some critical $i$ which is the same for all objects, then this $i$ must lie between 64$^\circ$ and 69$^\circ$ in order to accommodate both GS 2000+25 ($A=0.068$) and U Gem ($A=0.21$) in figure 2.  The cutoff of $68^\circ$, used in figure 3, separates almost all objects with $q<0.65$ into those below the bound of figure 3 and those above it, suggesting a sudden change in $A$ near this $i$.  This value is consistent with the rather small range allowed by figure 2.  In addition, figure 3a suggests that $A$ increases around $q\approx 0.65$, consistent with the large $A$-values of EM Cyg and BT Mon in table 2.  We suggest that this change in $A$ at large $q$ may result from fundamentally different asymmetries present in systems in which accretion is not driven by Roche lobe overflow.  Roche lobe overflow is unstable in any system with $q>0.83$, and may be unstable in systems with smaller $q$ (FKR, pp 56 ff).  Finally, both 4.1 and 4.2 suggest that the effects of high-$i$ and high-$q$ combine when both are present: each of the two sections identifies one object with exceptionally high $A$, and the two objects thus identified both have high-$i$ and high-$q$.

As described in the discussion following equation 4, the value of the asymmetry parameter $A$ depends on a disk's dimensionless asymmetries.  Thus, if a population of objects has a constant value of $A$, it suggests a consistency in the asymmetries present in that population.  Though it would be possible for the same value of $A$ to originate from different asymmetries in different systems (for example, hot spots in some, eccentric orbits in others), it would be surprising for these distinct phenomena to give rise to the same value of $A$ in many objects.  Specifically, the constancy of $A$ in systems with $q<0.65$ and $i<68^\circ$ suggests that all accreting binaries with $q<0.65$ have disks with similar asymmetries.  The inclination effect cannot be ascribed to a physical difference in the underlying disks, since there is no reason to believe that the high-$i$ systems are physically different then their low-$i$ counterparts.  The existence of an inclination effect shows that the phase shift is not entirely dependent on the disk itself, but also on how we observe the disk.  In particular, the presence of an inclination effect is further evidence against traditional physical displacement models, in which the direction of gas motion in the line-emitting region is assumed to be the sole factor determining the phase shift, independent of inclination.

The effect of inclination on the observed phase shift casts a new light on the study of SW Sex objects, which are a subclass of eclipsing CVs (e.g. G01) and thus all have high inclination.  The large phase shifts of SW Sex systems have sometimes been explained by invoking physical displacement models (e.g. Dhillon et al. 1997, G01).  We suggest that phase shifts in SW Sex objects should, instead, be considered as a special case of the inclination effect on observed phase shifts.  Though many of the largest phase shifts in figure 3b are in SW sex objects, other high-inclination objects show comparable, sometimes larger, shifts.  Moreover, physical displacement models--in addition to the problems discussed in section 2.1--predict a low-inclination analog to SW Sex objects, with equally large phase shifts.  Figure 3 shows that no such objects exist in the low-$i$, low-$q$ regime.  Future work on phase shifts in these systems should thus consider explanations which could only appear in systems viewed at high inclination.  Within the framework of measurement offset models, this will almost certainly involve a strong, phase-dependant effect from self-absorption in the disk.  Since phase-dependant absorption features are among the defining characteristics of SW Sex systems (Dhillon et al. 1997, G01), such an explanation is plausible.

\section{Negative Phase Shifts: V2107 Oph and V838 Her}

We now consider the two objects excluded from figure 3, both of which show large, negative phase shifts.  Physical displacement models imply a direct relationship between the observed phase shift and the velocity of gas in the emitting region, requiring these two objects to have gas flows opposite in direction to all others.  In contrast, measurement offset models allow positive and negative phase shifts to come from otherwise similar objects.  In measurement offset models, negative phase shifts occur precisely when $\phi_\mathit{off}>0.5$.  As we discussed in section 3.1, the largest phase shifts, both positive and negative, occur when $\phi_\mathit{off}\sim 0.5$ and $K_\mathit{off}/K_1$ is large.  In particular, figure 1 shows that, for $K_\mathit{off}/K_1\sim 1$, there is a near discontinuity in the observed phase shift, which jumps very quickly from +0.25 to -0.25  when $\phi_\mathit{off}$ becomes larger than 0.5.  

The system parameters needed to produce large positive phase shifts, e.g. in EX Dra and GS 2000+25, are very similar to those which could produce the negative shifts in V2107 Oph and V838 Her.  All that is needed to explain the existence of large, negative, phase shifts is that the values of $\phi_\mathit{off}$ in different systems cover a range extending beyond 0.5; this is reasonable, since there is no reason to expect a hard cutoff at any particular value of $\phi_\mathit{off}$.  The systems showing negative phase shifts are precisely those with $\phi_\mathit{off}>0.5$, and these systems are clearly rare.  Thus, we expect the distribution of $\phi_\mathit{off}$ to extend only slightly beyond $0.5$, into precisely the range needed for large, negative phase shifts.  The other criterion for large, negative shifts is that $K_\mathit{off}/K_{1}$ is large, and both V2107 Oph and V838 Her lie in regimes in which this is expected: (1)V2107 Oph is a very low-$q$ object, which implies a large $K_\mathit{off}/K_{1}$ by equation 3, and (2) V838 Her is a high-inclination object with $q>0.65$, where we found the largest $A$-values in sections 4.2 and 4.3.  Again by equation 3, large values of $A$ imply large $K_\mathit{off}/K_{1}$.

Filippenko et al. (1997) used emission line radial velocities to derive a mass ratio of 0.01 for V2107 Oph, a value which they consider, probably rightly, to be unphysical.  We now show that this a natural result of our explanation of this object's large, negative phase shift.  The phase shift of V2107 is -0.24, very close the the minimum possible phase shift (-0.25) which can occur when $K_\mathit{off}/K_1\leq 1$.  Figure 1 shows that, for $K_\mathit{off}/K_1<<1$ such extreme phase shifts are impossible, while for $K_\mathit{off}/K_1>>1$, such small shifts require values of $\phi_\mathit{off}$ substantially larger than 0.5.  If, as above, we assume that $\phi_\mathit{off}$ is only slightly larger than 0.5 in V2107 Oph, we must also conclude that $K_\mathit{off}/K_1\approx 1$.  In this case, the observed radial velocity curve arises as the sum of two curves--$v_1(\phi)$ and $v_\mathit{off}(\phi)$--which are of comparable amplitude and nearly in antiphase, causing them to nearly cancel when added.  The amplitude of $v_{obs}$ is thus much smaller than the true amplitude of $v_1$, causing a spuriously small mass ratio to be measured.  This effect can easily produce $K_1/K_{obs}\geq 4$, causing the mass ratio measured using emission lines to be the same factor too small and explaining the value derived by Filippenko et al.  

V2107 Oph provides a intriguing case for measurement offset models, which simultaneously explain both its very negative phase shift and the unphysically extreme mass ratio determined from its emission lines.  In addition, our analysis points to the possible utility of measurement offset models for the interpretation of phase-shifted disk emission line RV curves: with knowledge of the offset amplitude and phasing, one can work back from an observed emission line radial velocity curve to determine the actual velocity of the primary and thus the true mass ratio.  Though we have not given a quantitative estimate of the true mass ratio of V2107 Oph, we have been able to explain, by considering its phase shift, the qualitative relationship between the true mass ratio and the value derived from emission lines.  The work presented here is not sufficient to do more than this, but we are optimistic that more quantitative results may eventually be possible by using reasoning analogous to our discussion of V2107 Oph.  This may ultimately allow robust measurements of mass ratios to be derived from emission line radial velocity curves, a possibility which we discuss in the next section.

\section{Future Work}

We have shown that measurement offset models are consistent with existing data and that they provide a framework for understanding some previously puzzling observations, including the large, negative phase shifts of V2107 Oph and V838 Her.  In addition, they provide a quantitative explanation of the $q-\phi_{obs}$ relationship, which generalizes the $P-\phi_{obs}$ relation of Mason et al. (2001).  We therefore believe that measurement offset models provide the correct framework for the analysis of disk emission line radial velocity curves.  We would ultimately like to have a detailed physical model of the phenomena producing the phase shift, and to be able to derive robust estimates of the mass ratio from disk emission line radial velocity curves; the adoption of measurement offset models is an essential step towards achieving both of these goals. 

In this section we describe the next steps towards both a physical model of the phase shift and a technique for  the robust determination of mass ratios from disk emission lines.  In section 6.1, we show that a robust determination of the mass ratio may be made using disk emission lines if the asymmetry parameter $A$ is known, and we discuss future observations which would test and refine the ideas presented in this paper, with the specific goal of allowing the robust determination of mass ratios from disk emission line radial velocity curves.  In section 6.2, we describe an approach to understanding the physical phenomena underlying the offset velocity and thus the phase shift.

\subsection{Robust Mass Determinations from Disk Emission line RV curves}

We now describe how the true mass ratio $q$ is constrained by a disk emission line radial velocity curve if $A$ is known.  If, as suggested in section 4, $A$ is constant for low-$q$, low-$i$ accreting binaries, then a value may be assumed in these objects, even when the true value can't be measured directly; the methods described here then allow a determination of $q$ from disk emission line radial velocity curves.  We do not assume that $K_{obs}$, the observed radial velocity semiamplitude, is equal to $K_1$, the maximum radial velocity of the primary.  Rather, we describe how to invert the method of section 4.2, where we derived $A$ from the observed quantities $\phi_{obs}$, $K_{obs}/K_2$, and $q$.  In this section, we assume $A$ and use $\phi_{obs}$ and $K_{obs}/K_2$, obtainable from radial velocity curves, to constrain $q$.  

In order to invert the method of section 4.2, we must determine which values of the true mass ratio $q$ are consistent with given values of $A$, $\phi_{obs}$, and $K_{obs}/K_2$.  An analytic approach to this problem leads to an iterative procedure which is not necessarily convergent.  However, the problem may easily be solved numerically by determining, for each $q_0$ within a range of plausible values, whether $q_0$ is consistent with the given values of $A$, $\phi_{obs}$, and $K_{obs}/K_2$.

In general, there will be at most two values of $q$ which are consistent with given values of $A$, $\phi_{obs}$, and $K_{obs}/K_2$, and each of the possible values of $q$ will have a (unique) corresponding value of $\phi_\mathit{off}$ which produces the observed $K_{obs}/K_2$.  If the uncertainties in $A$, $\phi_{obs}$, and $K_{obs}/K_2$ are all propagated through the above calculation, then each of the allowed values of $q$ is the center of some allowed range of $q$, with a corresponding allowed range of $\phi_\mathit{off}$.  In many cases, the allowed ranges of $q$ will overlap, and the presence of two solutions will simply increase the uncertainty in the derived value of $q$; in other cases, there may be two distinct ranges of $q$ allowed by the data.  In the latter case, identifying the correct range of $q$ requires an independent constraint on either $q$ or $\phi_\mathit{off}$; even in the former case, a constraint on $\phi_\mathit{off}$ will reduce the uncertainty in the measured $q$.  It is therefore of some interest to develop empirical constraints on the range of $\phi_\mathit{off}$ present in accreting binaries, in order to better constrain the values of $q$ determined by the method we have described here.  

In order to further test measurement offset models, verify the constancy of $A$, and to refine and extend the methods described above, we suggest an observing campaign to obtain disk emission line radial velocity curves of accreting binaries whose mass ratios are already well-determined by other means.  This would allow a direct determination of $A$ in more objects than just those in table 2, and thus a confirmation of its constancy and a better determination of its value and the uncertainty in that value.  Including objects with $60<i<75$ would allow a determination of the range in $i$ over which the canonical $A$ is valid, and the inclination at which self-absorption becomes significant and increases $A$.   There is likely to be some range of $i$ in which self-absorption may or may not be important, i.e. in which $A$ is sometimes consistent with the low-$i$ value and sometimes much larger; if this range of $i$ is small, it would suggest a standard opening angle for disks in accreting binaries, potentially providing a useful constraint on disk structure.  Finally, a study of emission line radial velocity curves of systems with well-determined $q$ would allow a study of the distribution of $\phi_\mathit{off}$, which would further improve the mass ratio determinations obtainable by future radial velocity studies.

\subsection{The Physical Origin of $v_\mathit{off}$}

As mentioned in section 2.2, there are many  possible physical phenomena which could produce a measurement offset and thus a phase shift.  In order to distinguish between the various possibilities---hot spots, eccentric gas orbits, warped disks, etc., it is necessary to know how each of these phenomena affects the observed emission line profile and how these changes in the profile translate into a measurement offset.  A full physical understanding of $A$ requires knowing what asymmetries actually appear in accretion disks, and how these asymmetries affect the measured disk emission line radial velocity curve.  In this section, we discuss two complementary approaches to this problem: (1)  analytic approximations to $v_\mathit{off}$ for various physical phenomena, and (2) interpretation of $A$ in terms of detailed models of disk structure and radiative transfer.  

\subsubsection{Physical interpretation of $v_\mathit{off}$ via analytic approximations}

There are, in the literature, several analytic relations between between various types of asymmetry and the measurement offsets they produce.  For example, S70 adopts the assumption that, for a disk with a hot spot, the measured velocity is equal to the intensity-weighted mean velocity of the entire disk, which will be shifted towards the velocity of the hot spot at all times.  Paczynski (1977, henceforth P77), considering the possible effect of tidal distortions on measured velocities, assumes that the velocity offset is equal to the average of the most extreme redshifted and blueshifted velocities at the outer edge of the disk; this analytic approximation to the measurement offset has appeared in a number of papers, e.g. Wu et al. (2001).  In principle, these relations allow the interpretation of an observed measurement offset in terms of the physical structure of the disk, but these approximations have never been carefully tested and thus must be used with care.  

In order to have confidence in analytic approximations to the measurement offset, it is essential that these approximations be tested using model profiles representing disks with various asymmetries.  Even for a simple elliptical disk with no hot spot, it is easily shown that the intensity-weighted mean of S70 gives a smaller estimate of the measurement offset than the estimate given by P77; it is thus unclear which, if either, of the models gives the correct estimate of $v_\mathit{off}$.  Moreover, there are many methods of measuring the radial velocity of a disk from its emission lines, and these methods often give different measurements of $v_{obs}$ and thus of $v_\mathit{off}$ (e.g. Orosz et al. 1994, Shafter 1983).  Thus, even if an analytical approximation to the offset velocity is reliable for one method of measuring the radial velocity, it may not be for another.  In order to use analytic approximations like those employed by S70 and P77, it is therefore necessary to understand the relationship between various physical asymmetries, the methods employed to measure radial velocities, and the resulting measurement offsets; this can be achieved only by applying common measurement methods to model profiles, in which the underlying asymmetries and true radial velocity are known, and $v_\mathit{off}$ can therefore be computed directly. 

Finally, we note one limitation to analytic approximations to $v_\mathit{off}$: even if they are tested against model profiles and shown to be robust, they do not provide an immediate answer to the question of what causes the phase shift.  In order to use these approximations, one must first assume that the physical phenomenon causing $v_\mathit{off}$ is known.  For example, if one uses the approximation given by P77 to derive the velocity field required to produce a given $v_\mathit{off}$, one is implicitly assuming that the velocity field is the only feature contributing to $v_\mathit{off}$.  These analytic approximations may nonetheless be useful.  For example, the intensity of hot spots in many disks is well-established by doppler tomography, and it would be worth knowing what fraction of $v_\mathit{off}$ may be attributed to the hot spot alone.  In addition, it is possible that these analytic approximations may be useful in demonstrating that certain types of asymmetry make a minimal contribution to $v_\mathit{off}$, providing a useful constraint on the origin of $v_\mathit{off}$.  We conclude that a careful treatment of these analytic approximations is worthwhile, and may provide useful guidelines to what types of asymmetries are relevant to $v_\mathit{off}$.  

\subsubsection{$A$ as a new constraint on models of accretion disks}

Analytic approximations like those discussed in the previous section may provide some insight into the general origins of $v_\mathit{off}$ and thus $A$, but a detailed understanding of the behavior of $A$ will most likely come only from full disk models.  $A$ provides a new observational constraint on disk models, and if models are consistent with this constraint, they will give a complete description of the physical phenomena which contribute to $A$.  In this section, we discuss the constraint which $A$ places on disk models, and suggest how the results of this paper might be used to better understand disk structure.  

The constraint on disk structure provided by knowing $A$ is complementary to the constraint provided by doppler tomography.  Although doppler tomography is very good at identifying intensity variations in the disk's surface, it cannot reliably identify a velocity anisotropy: unless the true velocity of the primary is known, it is impossible to determine the degree of velocity asymmetry present in a doppler tomogram.  Indeed, a perfectly circular doppler tomogram may arise from an asymmetric velocity field, as was shown by Marsh, Horne, \& Shipman (1987); the velocity asymmetry is detectable in the doppler tomogram only if $v_1$ is known.  Such velocity anisotropies, however, contribute to the asymmetry parameter $A$, which we have constrained indirectly even in situations where $v_1$ is not known.  

In order to use $A$ as a constraint on disk models, it is necessary to determine $v_\mathit{off}(\phi)$; this requires producing model emission line profiles as a function of viewing angle, then determining $v_{obs}(\phi)$ by applying standard radial velocity measurement methods to the model profiles.  Since computing $A$ requires producing model profiles, it is possible that a full understanding of $A$ will require both 3D, SPH disk models, and a solution to the radiative transfer equations to produce realistic profiles.  In particular, understanding high-$i$ systems, in which self-absorption by the disk is significant, will almost certainly require a full solution of the radiative transfer equations.  We speculate that in the low-$q$, low-$i$ regime in which $A$ is constant, a simple treatment of radiative transfer will suffice: if self-absorption by the disk were relevant at low inclinations, we would expect to see an inclination effect in low-$i$ systems, but there is no evidence for such an effect in figure 2 or figure 3.  

We suggest the following approach to using $A$ as a constraint on disk models: (1) Using the best available 3D, SPH accretion disk models, compute $A$ for a variety of $q$-values up to $q=0.65$, adopting a simplified radiative transfer model neglecting self-absorption.  These models should produce a constant value of $A\approx 0.07$, independent of $q$.  If they do, this would provide empirical support for both the underlying disk model and the assumption that self-absorption by the disk is negligible in this regime.  (2) In order to be consistent with all the observations discussed in this paper, disk models of any $q<0.65$ must have $A=0.07$ when viewed at low-inclination, and larger $A$ when viewed at high inclination.  Thus, once models have been shown to be successful in the low-$i$ regime, the same disks should be tested in the high-$i$ regime, making no change to the disk, but introducing realistic radiative transfer in the production of the model profile.  If the models are correct, this should be sufficient to reproduce the change in $A$ at high $i$.  We note that this places a fully 3D constraint on disk models, by constraining both the angle at which self-absorption becomes important, and the effect it has on the observed profile.  (3) The high-$q$ regime, in which accretion may not be driven by Roche lobe overflow, may require a separate class of models.  It would certainly be interesting if models of disks driven by roche lobe overflow show a sudden change in $A$ near $q\approx 0.65$, but we suspect that explaining the $q$-depedence of $A$ will require invoking alternate modes of accretion.  Though modeling alternate modes of accretion may present a substantial new challenge, parts (1) and (2) of this program should be feasible with existing code.  They would provide a new observational check on fully three-dimensional disk models and possibly shed light on the physical origin of the phase shift.

\section{Conclusions}
The presence of phase shifts in radial velocity curves measured using disk emission lines has long prevented the use of this data for trustworthy determinations of the mass ratio of compact accreting systems; given that the lines have a different phasing from the motion of the primary, the semiamplitude of RV curves obtained from emission lines may not reflect the motion of the primary.  The relationship between the observed disk emission line RV semiamplitude and the true semiamplitude of the primary's RV depends on the model one adopts for the phase shift.  In order to reliably interpret disk emission line RV curves it is therefore essential to have a trustworthy model for the origin of the phase shift.  We have considered two possible scenarios, which we call ``physical displacement models" and ``measurement offset models."  Our conclusions are as follows:
\begin{enumerate}
\item Any possible explanation of the phase shift must invoke either a physical displacement model, a measurement offset model, or both; though physical displacement models are generally adopted in the literature, they are contradicted by observed line widths, line profiles, and eclipse observations.  Measurement offset models, however, are consistent with existing data.

\item Measurement offset models predict an upper bound on the phase shift, which depends on the mass ratio and the ``asymmetry parameter" $A$, a parameter we define to quantify the total degree of the dimensionless asymmetries in a disk.  This upper bound is apparent in phase shift data drawn from throughout the literature.

\item We find that $A=0.068$ is consistent for $q\lesssim 0.65$ and $i\lesssim 68^\circ$, but that $A$ is larger in high-$i$, high-$q$ systems.  We derive this result using two logically independent methods; the consistency between these two methods gives further support for measurement offset models.  We speculate that the dependancy of $A$ on $i$ and $q$ reflects the effect of self-absorption on the observed line profile of high-$i$ systems, and a change in disk structure in systems which are not driven by Roche lobe overflow.  

\item The previously puzzling negative phase shifts of V838 Her and V2107 Oph are naturally explained by measurement offset models, as is the unphysically small mass ratio derived for V2107 Oph using disk emission lines.  Our analysis of this object suggests how measurement offset models may be used to interpret observed emission line RV curves in terms of the actual RV of the primary.

\item Measurement offset models provide a useful framework of the analysis and interpretation of RV curves from disk emission lines.  Adopting these models implies a subtle relationship between the observed disk emission line RV semiamplitude and the true RV of the primary: small phase shifts may go along with highly spurious emission line semiamplitudes in some cases, but may be reliable in others.  

\item We have shown that our results, if confirmed, could lead to a new technique for a robust determination of mass ratios from disk emission line radial velocity curves.  Towards this end, we outline an observational program which would test and refine the conclusions of this paper.  

\item The results of this paper provide a new constraint on disk models.  This constraint is complementary to existing ones, and includes a constraint on the 3D structure of accretion disks.  In section 6.2, we discuss how our work can be used to test disk models, and how these models might give insight into the physical origin of $A$.

\end{enumerate}

\acknowledgements 
This work was supported by NSF Graduate Research Fellowship DGE-0202738 to AGC and NSF/AST
grant 0407063 to CDB.  The authors thank A.M. Hughes and M. Buxton for their comments.  AGC  thanks M. Eracleous and A.M. Hughes for their enthusiasm and interest, and M. Garcia for his encouragement.  We also thank  our anonymous referee for a helpful report.

\end{document}